\documentclass[aps,preprintnumbers,prl,twocolumn,superscriptaddress]{revtex4}
\usepackage{epsfig,latexsym,cancel,amssymb,amsmath}
\usepackage{graphicx}
\usepackage{epstopdf}
\usepackage{hyperref, graphicx, slashed, xcolor, amsmath}

\allowdisplaybreaks

\newcommand{\ta}{\text}
\newcommand{\tr}[1]{\text{#1}}
\newcommand{\eV}{\text{eV}}
\newcommand{\GeV}{\text{GeV}}
\newcommand{\brr}[1]{\left(#1\right)}

\graphicspath{{./figures/}}
\usepackage{xcolor}
\usepackage{xspace}

\begin{document}

\title{Radio Signals from Axion Star-Neutron Star Binaries} 

\author{Chris Kouvaris}
\affiliation{Physics Division, National Technical University of Athens, 15780 Zografou Campus, Athens, Greece}
\author{Tao Liu}
\affiliation{Department of Physics, the Hong Kong University of Science and Technology, Clear Water Bay, Kowloon, Hong Kong S.A.R., P.R.C.}
\author{Kun-Feng Lyu} 
\affiliation{School of Physics and Astronomy, University of Minnesota, Minneapolis, MN 55455, U.S.A.}
\affiliation{Department of Physics, the Hong Kong University of Science and Technology, Clear Water Bay, Kowloon, Hong Kong S.A.R., P.R.C.}

\begin{abstract}

Axion stars could form binaries with neutron stars. Given the extremely strong external magnetic field exhibited by individual neutron stars, there can be a substantial conversion of axions to photons in these binaries. The photon emission is doubly modulated  due to the neutron star spinning and the axion star orbiting, yielding a  unique discovery signal. Similar features are also generated in binaries between a neutron star and an axion-clouded black hole. Encouragingly, such binaries are found to be within the reach of ongoing and upcoming experiments (e.g., the Five hundred meter Aperture Spherical Telescope and the future Square Kilometer Array) for certain parameter regions. They thus provide a promising astronomical laboratory for detecting axions and axion dark matter.
\end{abstract}

\maketitle

\section{introduction}
Axions are originally proposed  to solve strong CP problem in QCD~\cite{Peccei:1977hh,Peccei:1977ur,Weinberg:1977ma,Wilczek:1977pj,Abbott:1982af,Preskill:1982cy,Dine:1982ah}, while axion-like particles (ALPs) are generically predicted by string theory~\cite{Arvanitaki:2009fg,Svrcek:2006yi,Cicoli:2012sz}, as a partner of geometric moduli and dilaton. Axions and ALPs (below we universally denote them as ``axions'' for simplicity) can both serve as a primary component of dark matter (DM) in the Universe, with a mass range spanning   more than twenty orders of magnitude~\cite{Marsh:2015xka}.  Because of their significant role in fundamental physics, axions have been well-studied in various contexts so far (for a recent review, see, e.g.,~\cite{Irastorza:2018dyq}).

The axions interact with the Standard Model (SM) of particle physics feebly, with a strength generically suppressed by their decay constant. The detections of axions so far rely mainly on their electromagnetic (EM)  coupling  
\begin{equation}
\mathcal{L}  \supset - c_\gamma \dfrac{\alpha_{e}}{4 \pi f_a} a F_{\mu\nu} \tilde{F}^{\mu\nu} \ .
\end{equation}
Here $a$ is the axion field, $m_a$ and $f_a$ are its mass and decay constant respectively, $\alpha_e$ is the fine structure constant, $F_{\mu\nu}$ is the EM field strength, and $\tilde F_{\mu\nu}$ is its Hodge dual.  The dimensionless coupling $c_\gamma$ is model-dependent. Its value can vary from $\sim \mathcal O(1)$ to many orders of magnitude higher~\cite{Choi:2015fiu,Kaplan:2015fuy,Long:2018nsl,Agrawal:2017cmd}. In this context, a detection signal can be produced either in static magnetic field, where a flux of axions are converted to photons via Primakoff process, or in a macroscopic axion field, where cosmological birefringence may occur when linearly polarized light travels across this field~\cite{Carroll:1989vb}. 

The extreme environment of stars is especially suitable for axion production and conversion. For example, axions can be produced in solar plasma or in core-collapsing supernovae such as SN1987A from thermal photons. These axions are then converted back to photons in an artificial or astrophysical magnetic field. The observations from the CERN Axion Solar Telescope and the Solar Maximum Mission satellite have set two of the most important constraints for the axion EM coupling~\cite{CAST:2017uph,CAST:2004gzq,CAST:2007jps,Rapley17:new}. In parallel with this, the polarization measurements of Cosmic Microwave Background~\cite{Harari:1992ea,Lepora:1998ix,Lue:1998mq}
and pulsar light (with either individual pulsars~\cite{Liu:2019brz,Caputo:2019tms} or pulsar polarization arrays~\cite{Liu:2021zlt}) provide a sensitive way to detect cosmological/astronomical axion field in certain mass regions.

Separately, in an axion-rich environment (e.g., in the early Universe or near black holes (BH)), macroscopic objects such as axion miniclusters~\cite{Fairbairn:2017sil,Hogan:1988mp,Visinelli:2018wza,Eggemeier:2019khm,Tkachev:2014dpa}, axion clumps~\cite{Kolb:1994fi,Hertzberg:2020dbk} and axion stars (AS)~\cite{Kolb:1993zz} can form, due to the effects of density fluctuations and self gravity (for recent reviews, see, e.g.,~\cite{Zhang:2018slz,Eby:2019ntd}). While traveling in space, these objects may collide with neutron stars (NS) and even magnetars, and then be substantially converted to detectable radio signals in their magnetic fields which are known to be extremely strong. 
This scenario has attracted a lot of attentions in last years~\cite{Iwazaki:2014wka,Raby:2016deh,Pshirkov:2016bjr,Bai:2017feq,Buckley:2020fmh,Bai:2021nrs,Nurmi:2021xds,Wang:2021hfb}.  
Furthermore the AS can potentially form binaries with the NS either by fragmentation which pairs the AS and supermassive progenitor of the NS from the start, or by later-time gravitational capture. 
Then as the AS is subjected to the NS strong magnetic field in such a binary, the axions can be converted to photons in large number. The generated signals have double modulations with time due to the spinning of the NS and the Keplerian orbiting of the AS around its companion, thus yielding a signature which can unambiguously distinguish this binary from the collision events and other possible astrophysical sources of radio frequencies.

Alternatively,  the axions can form dense clouds surrounding a BH. ``Gravitational atoms'' (GA) are such an example~\cite{Brill:1972xj,Detweiler:1980uk}.  
If the superradiance instability occurs near a spinning BH, the axion occupation number for certain states can grow exponentially by extracting energy and angular momentum from the BH~\cite{Brill:1972xj,Detweiler:1980uk,Dolan:2007mj,Yoshino:2013ofa,Brito:2015oca,Endlich:2016jgc,East:2018glu}, yielding  strong accumulation concentrated at, e.g., the ``Bohr radius'' of this BH. Some other formation mechanisms also exist for the axion-clouded BH (ABH) other than the superradiance instability. For example,  the axions if playing the role of DM could be part of an accretion disk around a BH~\cite{Jacobson:1999vr,Wong:2019yoc,Clough:2019jpm,Hui:2019aqm}. Also, the axions could inhabit supermassive stars that after a supernova might not be carried away from the blast, thus forming a cloud around the newly born NS or BH (see, e.g.,~\cite{Kouvaris:2010vv}).

In this letter we will examine three types of binaries, where the NS is accompanied by a (i) diluted AS (denoted as ``AS1''), (ii) dense AS (denoted as ``AS2'') and (iii)  axion-clouded BH (denoted as ``ABH''), respectively. We will demonstrate that the doubly-modulating signals arising from these scenarios could be detected by the ongoing and upcoming experiments such as the Five hundred meter Aperture Spherical Telescope (FAST)~\cite{Zhu:FAST,Nan:2011um} and the future Square Kilometer Array (SKA)~\cite{ska,Weltman:2018zrl}.

\section{Binary Profile}

The diluted AS form where the axion number density is relatively small. They typically have a mass~\cite{Braaten:2015eeu}
\begin{equation}\label{eq:dilu_M}
M_\text{AS1} \lesssim 1.0 \times 10^{-8} M_\odot \brr{\dfrac{f_a}{10^{14}\GeV}} \brr{\dfrac{10^{-6}\eV}{m_a}} \ ,
\end{equation}
and a radius
\begin{equation}\label{eq:dilu_R}
R_\ta{AS1} \approx 2.7{\rm km} \brr{\dfrac{10^{-8} M_\odot}{M_\ta{AS1}}} \brr{\dfrac{10^{-6}\eV}{m_a}}^2  \ .
\end{equation}

In the dense case, the occupation number of axions is so big that their self-interactions, which are usually attractive, can not be neglected. With a Thomas approximation, the radius of dense AS is given by~\cite{Braaten:2015eeu}
\begin{eqnarray}
R_\ta{AS2} &=& 0.95\ta{m} \\ 
&& \times \brr{\dfrac{10^{14}\GeV}{f_a}}^{\frac{1}{2}}\brr{\dfrac{10^{-6}\eV}{m_a}}^{\frac{1}{2}}   \brr{\dfrac{M_\ta{AS2}}{10^{-12} M_\odot}}^{0.3} \ .  \nonumber
\end{eqnarray}
The dense AS can be viewed as oscillons~\cite{Visinelli:2017ooc}. The lifetime of QCD axion oscillons is $\sim 10^{3} m_a^{-1}$~\cite{Gleiser:1993pt,Salmi:2012ta,Fodor:2006zs}. 
The AS with such a short lifetime may have little astrophysical and cosmological effects, if they are produced in the early Universe. However, relatively flat axion potentials may yield long-lived oscillons~\cite{Zhang:2020bec}. In this study we will assume the existence of such long-lived AS and a certain amount of them thus can survive until today.

As for the ABH, let us consider the GA as an example. The gravitational fine-structure constant $\alpha_G$ and Bohr radius $r_G$ are defined  in terms of the BH and axion masses as~\cite{Baumann:2019ztm,Baumann:2019eav}
\begin{eqnarray}
\alpha_G &=& G M_{\rm BH} m_a = 0.075 \brr{\dfrac{M_{\rm BH}}{10^{-5} M_\odot   }} \brr{\dfrac{m_a}{10^{-6}\tr{eV}}} \ , \nonumber \\
r_G &=& \frac{1}{m_a \alpha_G} = 2600{\rm km} \brr{\dfrac{0.075}{\alpha_G}}  \brr{ \dfrac{10^{-12} \ta{eV}}{m_a} }    \ .
\end{eqnarray}
The superradiance instability requires $\alpha_G$ to be smaller than $\mathcal{O}(1)$. In the limit of $\alpha_G<<1$, the energy eigenstates can be approximated as $\langle t,\vec{r} |nlm\rangle = R_{nl}(r) Y_{lm}(\theta,\phi) e^{-i(\omega_{nlm} - m_a) t}$, where $\{n,l,m\}$ denote the principal, orbital and magnetic quantum numbers of the GA  respectively. 
The leading growing mode is $|211\rangle$ and the next-leading one is $|322\rangle$~\cite{Detweiler:1980uk}. The total mass for a fully-occupied $|211\rangle$ state ($M_{211}$) is subjected to two constraints here. The first one is the
saturation condition, i.e. the angular momentum extracted from the spinning BH can not be larger than its initial value, which is written (normalized by $G M_{\rm BH}^2$) as $\tilde{a}_i =J_i/(G M_{\rm BH}^2) \leq 1$. Therefore $M_{211}$ needs to be smaller than~\cite{Baumann:2019ztm} 
\begin{eqnarray}
M_\text{sat} &\sim & \alpha_G (\tilde{a}_{i} - 4\alpha_G) M_{\rm BH} \\
&\approx& 1.5 \times 10^{24} {\rm kg}   \brr{\dfrac{M_{\rm BH}}{10^{-5} M_\odot }}^2 \brr{\dfrac{m_a}{10^{-6}\eV}}  \ . \nonumber 
\end{eqnarray}
The second constraint comes from the bosenova limit~\cite{Yoshino:2012kn,Amin:2021tnq}, where the maximal occupation number of axions ($N_{211}$) is set to be $\sim [10 f_a/(\alpha_G^{3/2} m_a)]^2$. So  $M_{211}$ is constrained to be smaller than 
\begin{eqnarray}
&& M_\text{bos} \sim N_{211}    m_a  \\
&\approx& 4.2 \times 10^{23} {\rm kg}  \brr{\dfrac{10^{-5} M_\odot }{M_{\rm BH}}}^3 \brr{\dfrac{10^{-6}\tr{eV}}{m_a}}^4  \brr{\dfrac{f_a}{10^{15}\tr{GeV}}}^2     .  \nonumber 
\end{eqnarray} 
The maximally allowed $M_{211}$ is thus determined by the minimum between $M_\text{sat}$ and $M_\text{bos}$, namely $M_{211} \leq \min \{M_\text{sat}, M_\text{bos}\}$.

\begin{figure}
\centering
\includegraphics[width=8cm]{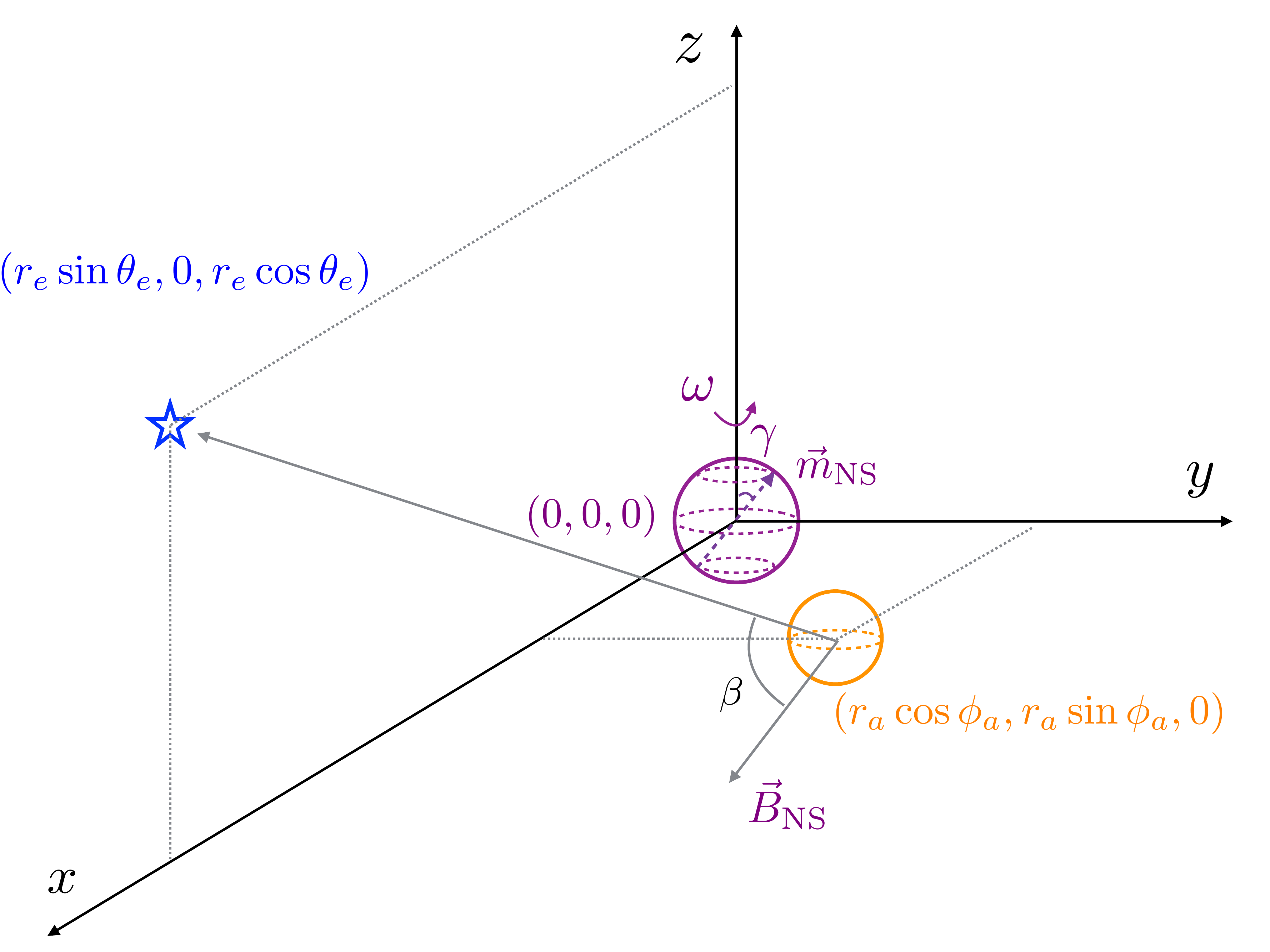}
\caption{Coordinate system used for describing the AS/ABH-NS binary. The purple and orange circles represent the NS and AS /ABH respectively, and the blue star represents the Earth. The NS spins around the $z-$axis with an angular velocity $\omega$.  Its magnetic dipole $\vec{m}_\tr{NS}$ is not aligned with its spin axis, subtending a misalignment angle $\gamma$. We have chosen this coordinate system such that the AS/ABH orbits the NS in its equator plane or the $x-y$ plane, and the Earth lies in the $x-z$ plane. 
}
\label{fig:config}
\end{figure}

The NS spin with a period ranging from a few milliseconds to seconds. Normally, the magnetic field on their surface is $\lesssim 10^{12}$G. But for magnetars, a class of extreme NS, their magnetic field can reach $\sim 10^{15}$G. For simplicity, we assume that the NS magnetic field is well approximated by a magnetic dipole $\vec{m}_\text{NS}$~\cite{Goldreich:1969sb} 
\begin{equation}
\vec{B}_{\rm NS}(\vec{r}) = - \dfrac{\mu_0}{4\pi} \dfrac{\vec{m}_\tr{NS}-3(\vec{m}_\text{NS} \cdot \hat{r})\hat{r}}{r^3} \ .
\end{equation}
This is justified so long as  the AS/ABH lies  within the light cylinder of the NS which has a radius $r_c=c/\omega\simeq 4.8 \times 10^4 T_s~{\rm s}^{-1}\text{km}$ ($T_s$ is the NS spin period). Outside the light cylinder, the toroidal magnetic field created by the NS magnetosphere  becomes dominant~\cite{Goldreich:1969sb}.

The coordinate system used to describe the AS/ABH-NS binary is shown in Fig.~\ref{fig:config}. We assume the separation between the two objects of the binary to be sufficiently big, to neglect curvature corrections. We assume $M_{\rm NS} \gg M_{\rm AS/ABH}$ also, such that the NS stays at rest to a good approximation, while the AS/ABH orbits the NS with  
\begin{equation}
r(\phi) = \dfrac{(1- e^2) r_0}{1+ e \cos\phi_a} \ , \quad r_0 = \dfrac{l_0^2}{G M_{\rm NS} (1-e^2) }   \ .
\end{equation}
Here $e$ and $r_0$ are the orbit eccentricity and semi-major axis respectively, $l_0$ is the AS/ABH angular momentum per unit mass, and $\phi_a$ is the azimuthal angle of the AS/ABH position (see Fig.~\ref{fig:config}).

Let us try to give a rough estimate for the existence of such binaries between AS and NS, following the corresponding estimate for the NS-NS binaries. There can be several mechanisms in play to form a NS-NS binary, ranging from direct gravitational capture, two-body tidal effects to three-body interactions. However,  the main mechanism which seems to agree with the NS-NS merger rate observed in gravitational interferometers is related to the fact that $\sim 70\%$ of stars form in binaries and out of them only $\sim 0.3\%$ are supermassive and will evolve into a NS or a BH after supernova explosion. Given the average stellar density of the Universe $\sim 3 \times 10^8M_{\odot}/\text{Mpc}^3$  and assuming the star formation rate for the last $10^{10}$ years to be stable, we arrive at the typical binary formation rate of $\sim 200 \text{yr}^{-1} \text{Gpc}^{-3}$ which is within the range expected for LIGO~\cite{Pian:2020vul}. Here the probability for a binary to have two supermassive progenitors is $\sim (3\times 10^{-3})^2$. 
The reason that most stars form in binaries is related to a fragmentation process. As molecular cloud collapses, its density increases while its Jeans mass reduces. The molecular cloud thus gets fragmented. We will invoke a similar mechanism for the formation of the AS-NS binary. We assume that  baryonic contraction takes place. That is, as baryons collapse, they draw into axions and AS as well. The Milky Way contains DM $\sim 10^{12}M_{\odot}$. Even if only $\sim 1\%$ of DM is in the AS and with a mass of $10^{-8}M_{\odot}$ (typical for the diluted AS), the AS population of the galaxy is $\sim 10^{18}$. Notably, the AS are distributed in the whole DM halo. Since we are interested in the AS-NS binaries and given that the NS lie in galactic disk, the relevant AS amount is $\sim 10^{18} V_{\text{disk}}/V_{\text{halo}}\sim 10^{15}$ (we use a typical  halo radius of 30 kpc and a disk radius of 15 kpc and height of 300 pc). Separately, there are $\sim 10^9$ NS in our galaxy. The majority of their supermassive progenitors were in binaries with other stars. Additionally, each of them has $\sim 10^{15-11}=10^4$ AS in the vicinity, given that the number of stars in our galaxy is $\sim 10^{11}$. Although some of these AS might escape or get absorbed by the nearby star, it seems quite probable that on average each supermassive progenitor and eventually NS might be in a bound state with at least one AS. The gravitational impact from the companion star on the AS motion could be neglected if the AS is close to the supermassive progenitor. Assuming again a constant rate for the NS formation, we can get an estimate on the AS-NS binary formation rate by dividing the total number of NS in the galaxy by the galaxy age ($\sim 10^{10}$ years) and $V_{\text{disk}}$, namely $\sim 5 \times 10^{-4}\text{yr}^{-1}\text{kpc}^{-3}$. With such a rate and due to an accumulative effect, as long as the AS-NS binaries are on average stable for more than a few thousand years, there will be several such binaries at the galactic disk per kpc$^3$ at any given time.

\section{Photon Emission from the Binaries}

The photon emission from the AS/ABH-NS binary is described by the axion-Maxwell equations, namely 
\begin{equation}
\partial_\mu F_a^{\mu\nu} = j_a^\nu = -  c_\gamma \dfrac{\alpha_{e}}{\pi f_a} \partial_\mu a \, \tilde{F}_{\rm NS}^{\mu\nu} \ .
\end{equation}
Here $\tilde{F}_{\rm NS}^{\mu\nu}$ is the dual of the NS EM field strength, and $F_a^{\mu\nu}$ is the EM field strength induced by axion current. Using the retarded Green function, one can find the vector potential near the Earth  
\begin{equation} 
A_a^\mu (\vec{r}_e, t) = \dfrac{1}{4\pi} \int d^3 r_a \dfrac{j^\mu_a(\vec{r}_a, t- |\vec{r}_e-\vec{r}_a|)}{|\vec{r}_e-\vec{r}_a|}   \ .
\end{equation}  
The radiation power is then given by
\begin{equation}\label{eq:radiation_power}
\begin{split}
\dfrac{dP}{d\Omega} =& \dfrac{1}{2} r^2 {\rm Re}[\hat{r}\cdot(\vec{E}_a\times\vec{B}_a^*)] = \dfrac{1}{2} r^2 m_a^2 |\vec{A}_a]^2 \sin^2 \beta \ ,
\end{split}
\end{equation}
where $\beta$ is the angle between $\vec{B}_{\rm NS} (\vec{r}_a)$ and the line of sight connecting the Earth with the AS/ABH.

Explicitly, for the three type of binaries we have 
\begin{eqnarray}
\dfrac{d P_i}{d \Omega}
& = & C_i \left( \dfrac{c_\gamma \alpha_e}{ f_a} \right)^2 |\vec{B}_{\rm NS}|^2 \sin^2\beta   \ , \label{eq:radp}
\end{eqnarray}
where $C_i$ is dimensionless, given by 
\begin{eqnarray}
C_{\rm AS1}
& = & \dfrac{M_\text{AS1}}{16 \pi^4 m_a} v^3   \ , 
\nonumber \\
C_{\rm AS2}& =&
4    \dfrac{M_{\rm AS2} m_a^2  R_{\rm AS2}^3}{\pi^3 (1+m_a^2 R^2_{\rm AS2})^4}  \ ,
\nonumber  \\
C_{\rm ABH} &=&
 \dfrac{1024 \alpha_G^7 M_{211}   }{  ( 4 + \alpha_G^2)^6   \pi^3 m_a} 
  \sin^2\theta_e  \ .   \label{eq:211_power}    \label{eq:Ceff}
 \end{eqnarray}
Here we have used the $1s$ wave function of Hydrogen atoms to approximate the profile of dense AS, following~\cite{Bai:2017feq}, and assumed the ABH to be in the leading mode $|211\rangle$. $\theta_e$ is the polar angle of the Earth position (see Fig.~\ref{fig:config}). The spectral flux density is then  derived by  
$S =  \frac{dP}{d\Omega} \brr{\mathcal{B} r_e^2}^{-1}$,
where $\mathcal{B} \sim 0.1 m_a/2\pi$ is a conservative estimate on signal bandwidth, based on Doppler effect.

\begin{table}
\resizebox{0.38\textwidth}{!}{
		\centering
		\begin{tabular}{ |c|c|c|c|c|c| }
		\hline
		$M_{\rm NS} (M_\odot)$ & $R_{\rm NS}$(km) & $B_{\rm NS}^0$ (G)  &  $\gamma$ (rad)   & $T_s$ (s) & $T_o$ (s) 	    \\	
		\hline 			
	   1.4 & 15 & $10^{12}$  &  0.3 & 1.0 & 10  \\	
					\hline
    $r_0$(km) &$\theta_e$ (rad) & $r_e$ (kpc) &  $ m_a$(eV) & $f_a$ (GeV) & $c_\gamma$  \\	
		\hline 			
   7.8$\times 10^3$ & $\pi/4$ & 1.0 & $10^{-6}$ &  $10^{14} (10^{15})$ & $10^5$  \\	
	   		\hline			
		\end{tabular}
		}
\resizebox{0.48\textwidth}{!}{
		\centering
		\begin{tabular}{|c|c|c|c|c|c|}
		\hline
		  $M_{\rm AS1} (M_\odot)$   & $R_{\rm AS1} $(km)  &  $M_{\rm AS2} (M_\odot)$  & $R_{\rm AS2} $(m) & $M_{\rm BH} (M_\odot)$  & $M_{211}$ (\text{kg})    \\	
		\hline 			
	    $10^{-8}$  &  2.7 & $10^{-12}$ & 0.95  &  $10^{-5}$ & $4.2 \times 10^{23}$    \\	
	   \hline
		\end{tabular}
		}
\resizebox{0.45\textwidth}{!}{
		\centering
		\begin{tabular}{ |c|c|c|c|c| }
		\hline
 $v(e=0)$ &  $v (e=0.5)$ &		$C_{\rm AS1}$ &  $C_{\rm AS2}$ & $C_{\rm ABH}$   \\	
		\hline 			
0.016  & 0.009 - 0.028 &$7.2 \times 10^{60} v^3$  & $ 4.7 \times 10^{55}$ & $1.3\times 10^{55}$   \\	
	   		\hline
		\end{tabular}
		}	
		\caption{Benchmark scenarios for the AS1-NS, AS2-NS and ABH-NS binaries. The upper subtable shows the common parameter values shared by all three scenarios (except $f_a$ which is taken to be $10^{14}(10^{15})$GeV for the AS1/AS2(ABH)-NS binaries), the middle one summarizes individual parameters specific to each scenario, and the bottom one gives the derived coefficients of  radiation power in these scenarios (see Eq.~(\ref{eq:Ceff})). $M_{\text{NS}}$, $R_{\text{NS}}$, and $B_{\text{NS}}^0$ are respectively the NS mass, radius and magnetic field at the NS surface. $T_o$ is the period of the AS/ABH revolution around the NS.}

		\label{tab:para}
\end{table}

To demonstrate these signals numerically, we will consider the benchmark scenarios defined in Tab.~\ref{tab:para}. We emphasize here that the plausibility of detection stands for orbital radii of $\mathcal{O}(10^3)-\mathcal {O}(10^4)$km (as is indicated by our choice of $r_0 = 7.8 \times 10^3$km for these scenarios).   For a binary, its Roche limit sets up the minimal separation between its two objects below which tidal forces will disrupt the relatively loose one. The Roche limit of, e.g., the AS1-BH binary~\footnote{The case for the ABH-NS binary is more involved. The appearance of a companion for the ABH changes the gravitational potential for axions, forming a ``gravitational molecule'' generically. In this case, the axion cloud is expected to evolve at the hybrid orbitals of the binary instead~\cite{Liu:2021llm}, as it happens at microscopic level to electron cloud in a chemical molecule. Its profile relies on both of the initial conditions of the binary (adiabatic versus diabatic) and its orbital hybridization~\cite{Liu:2021llm}. In this study, we will simply assume that the axion-cloud profile is not significantly deformed by gravitational effects of the companion, such that the GA with a companion is still a good approximation for the profile of the ABH-NS binary.}, is given by  
\begin{eqnarray}
&& R_{c,{\rm AS1}} = R_\ta{AS1} \brr{\dfrac{2 M_\ta{NS}}{M_\ta{AS1}}}^{\frac{1}{3}} \\ 
&\approx& 1800 \ta{km} \brr{\frac{M_{\rm NS}}{1.4 M_\odot}}^{\frac{1}{3}} \brr{\frac{10^{-6}{\rm eV}}{m_a}}^{\frac{2}{3}} \brr{\frac{10^{14}{\rm GeV}}{f_a}}^{\frac{4}{3}}  .  \nonumber
\end{eqnarray}
This limit is several times smaller than the $r_0$ benchmark value. For lower $f_a$, the diluted AS becomes larger and $R_{c,{\rm AS1}}$ also increases. 
Roche limit can be softly violated in reality. In that case although the loose object (the AS1 here) starts to lose mass to the NS to form a halo, its significant portion may remain intact and keep orbiting around the NS. But, as the tidal effects become  strong, this loose object will get disintegrated finally.

The plasma in the NS magnetosphere could affect photon emission in multiple ways. As mentioned earlier, the co-rotating plasma can create a toroidal magnetic field not ignorable outside the light cylinder. Moreover, the plasma deforms dispersion relation of photons. The generated effective mass for photons, namely  
\begin{eqnarray}
m_\gamma \sim w_p = \Big (\frac{4\pi  \alpha_e n_e }{m_e}\Big)^{1/2} \ ,
\end{eqnarray}
may have significant impacts on the radiation power calculations. Here $w_p$ is the plasma frequency, $m_e$ is the electron mass and $n_e$ is the  electron number density in the magnetosphere.  In the Goldreich-Julian model~\cite{Goldreich:1969sb}, $n_e$ is estimated as
\begin{eqnarray}
n_e \sim 7 \times 10^{-2} \frac{B_{NS}(r_a)}{1 {\rm G}}\frac{1 {\rm s}}{T_s} {\rm cm}^{-3} \ .
\end{eqnarray}
If $m_\gamma$ is comparable to or bigger than $m_a$, the axion-photon conversion will be enhanced or impeded~\cite{Pshirkov:2007st,Huang:2018lxq,Hook:2018iia}. This plasma effect can not be ignored in the calculations then. In the benchmark scenarios, however, the effective photon mass is given by $m_\gamma \sim 10^{-9}$eV, safely below the axion mass $m_a = 10^{-6}$eV. So we do not expect that our calculations are seriously affected by these plasma effects.

\begin{figure}[ht]
\centering
\includegraphics[width=9cm]{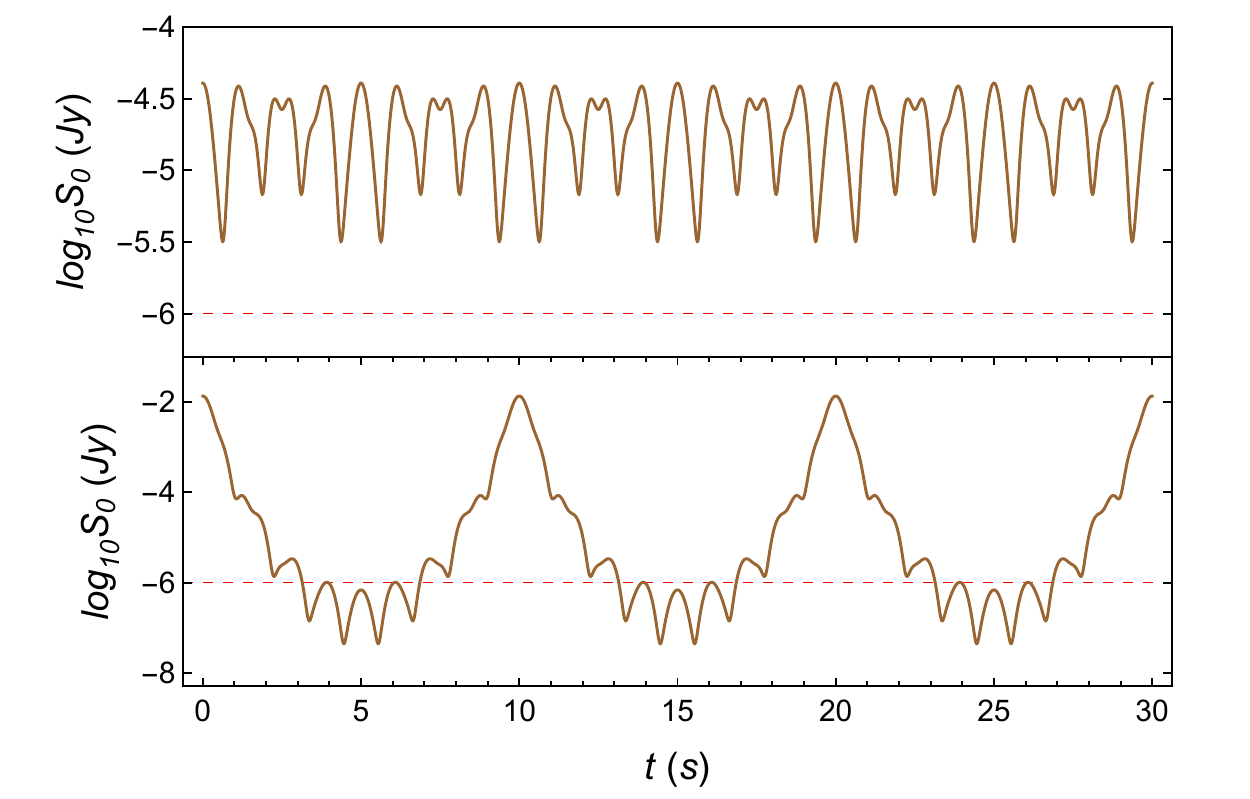}
\caption{Doubly modulated signals for the AS1-NS benchmark binary with $e=0$ (upper) and $0.5$ (bottom). The red-dashed line represents a baseline sensitivity at the FAST and SKA, with one-hour observation.}
\label{fig:flux}
\end{figure}

The three types of binaries share a similar profile of radiation power (see E.q.~(\ref{eq:radp})). Given the signals for the $i$-th type of binary, the signals for the other two can be obtained by multiplying a scaling factor $C_j/C_i$. The $C_{i,j}$ values in the benchmark scenarios are summarized in Tab.~\ref{tab:para}. For demonstration, we show the spectral flux density for the AS1-NS benchmark binary ($S_0$) in Fig.~\ref{fig:flux}. Due to the misalignment between the NS spin axis and its magnetic dipole, the magnetic field at the AS/ABH orbit changes periodically. The generated signals are thus modulated by the NS spinning and the AS1 Keplerian orbiting, at relatively small and large time scales respectively. 
For circular orbits, the $S_0$ envelope is approximately flat on the top and described by a trigonometric function at the bottom. This behavior originates from the convolution of the AS1 position vector and the magnetic field direction. The NS spin adds a higher frequency modulation due to its short period of $1s$.
For elliptic orbits, the modulation is more pronounced because the AS1-NS separation changes periodically. As is expected, a larger orbit eccentricity results in a stronger modulation. When the AS1 gets close to its perihelion, the signals can be enhanced by several orders of magnitude as the luminosity scales with the distance as $|\vec{B}_{\rm NS}|^2\sim r^{-6}$.

The AS1-NS radio signals could be detected by the FAST and SKA. The FAST and SKA have a frequency coverage ranging from 70 MHz to 3GHz, and from 50 MHz to 350 MHz (SKA-low) and 350 MHz to 14 GHz (SKA-mid), respectively, with an estimated sensitivity $\lesssim O(1)\mu$Jy $(\tr{hr}/t_\tr{obs})^{1/2}$~\cite{Nan:2011um,ska}. Here $t_\tr{obs}$ denotes observation time. Encouragingly, the spectral flux density presented in Fig.~\ref{fig:flux} lies above ($e=0$) or mostly above ($e=0.5$) the threshold, within an hour of observation. The doubly-modulated signal patterns thus could be established. We expect to achieve comparable sensitivities also for detecting the AS2-NS and ABH-NS benchmark scenarios.    

\begin{figure}[ht]
\centering
\includegraphics[width=8cm]{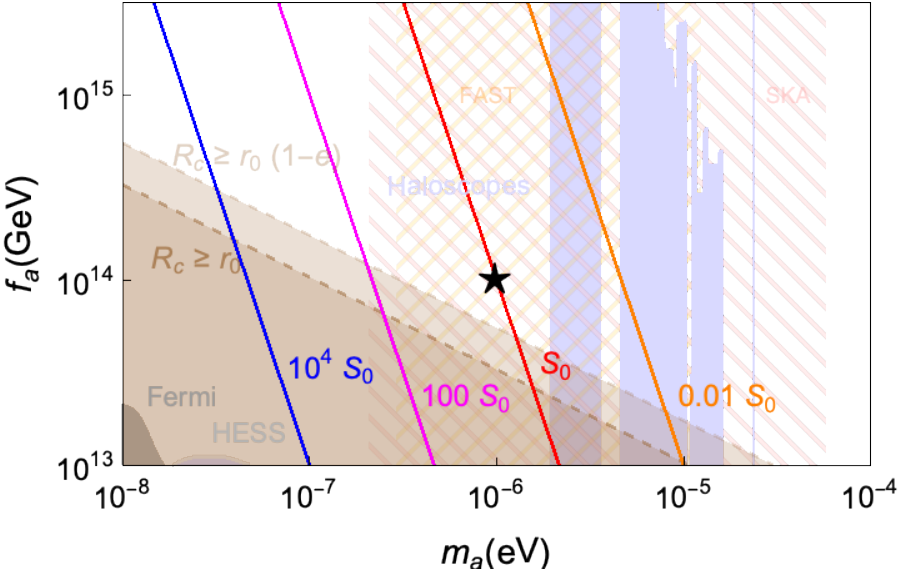}
\caption{Spectral flux density for the AS1-NS photon emission. The black star represents the benchmark scenario defined in Tab.~\ref{tab:para}.  The four different colored lines represent parameter points of constant signal strength as indicated. The light-blue and grey regions have been excluded by the haloscope and FERMI/HESS observations, respectively~\cite{ParticleDataGroup:2020ssz}. The shaded brown and light-brown regions are characterized by a Roche limit larger than the orbit radius ($e=0$) and the perihelion ($e=0.5$), respectively. The hatched regions denote the frequency coverage of the FAST (light-orange) and  the SKA (light-yellow). }
\label{fig:overall}
\end{figure}

Finally, we illustrate the potential signals of an AS1-NS binary for generic sets of parameters in Fig.~\ref{fig:overall}. According to the discussions above, the spectral flux density scales with $m_a$ and $f_a$ as:
\begin{eqnarray} 
S = \brr{\frac{10^{-6} {\rm eV} }{m_a}}^3 \brr{\frac{10^{14} {\rm GeV}}{f_a}} S_0   \ .
\end{eqnarray}
As $m_a$ decreases, the spectral flux density is enhanced quickly. 
This feature is demonstrated in Fig.~\ref{fig:overall} clearly. Notably, the low $m_a$ region ($i.e.$, $m_a \lesssim \mathcal O(10^{-7})$eV) is difficult for terrestrial radio observations, due to limitations such as ionospheric distortions and radio frequency interference of artificial signals. However, it could be well-probed by  future space- or lunar-based ultra-low-frequency radio telescopes, whose  frequency coverage is expected to reach a sub-Hertz level, several orders of magnitude below the low-frequency threshold of the FAST and SKA (see, e.g.,~\cite{BENTUM2020856}).

\section{Conclusion}

In this letter we have studied the potential radio signals produced in the AS/ABH-NS binaries. In such systems, the extremely strong magnetic field of the NS can convert with a significant rate axions to photons. Due to the spinning of the NS as well as the Keplerian orbiting of the AS/ABH around the former, the radio-signal intensity is doubly modulated with time, yielding a striking discovery signature for the existence of the AS or the ABH in our galaxy. These signals also provide an astrophysical probe for axions which could shed light on the nature of DM. At last, we would stress that our proposal based on this study strongly responds to the current intensification of the observational efforts on radio frequencies and the upcoming of a new era for radio astronomy.

\section{Acknowledgement}

K.-F. Lyu would like to thank Xi Tong, Qianhang Ding, Hongyi Zhang for useful discussions. T.~Liu is supported by the Collaborative Research Fund under Grant No. C6017-20G which is issued by the Research Grants Council of Hong Kong S.A.R.  KFL is partially supported by the DOE grant  DE-SC0022345.

\bibliography{references}
\end{document}